# Three-Dimensional Smoothed Particle Hydrodynamics Simulation for Liquid Droplet with Surface Tension

Hanifa Terissa[1,2], Agra Barecasco[1,2], Christian Fredy Naa[2]

[1]Graduate School of Natural Science and Technology, Kanazawa University, Kakuma, Kanazawa 920-1192 Japan,
[2]Faculty of Mathematics and Natural Sciences, Institut Teknologi Bandung, Jl. Ganesha 10, Bandung 40132 Indonesia,
E-mail: hanifa.t@students.itb.ac.id, agra.barecasco@s.itb.ac.id, chris@cphys.fi.itb.ac.id

**Abstract.** *We provide a basic method of Smoothed Particle Hydrodynamics (SPH) to simulate liquid droplet with surface tension in three dimensions. Liquid droplet is a simple case for surface tension modeling. Surface tension works only on fluid surface. In SPH method, we simply apply the surface tension on the boundary particles of liquid. The particle on the 3D boundary was detected dynamically using Free-Surface Detection algorithm. The normal vector and curvature of the boundary surface were calculated simultaneously with 3D boundary surface reconstruction using Moving Least-Squares (MLS) method. Before the reconstruction, the coordinate system was transformed into a local coordinate system. Afterwards, the surface tension force which depends on curvature of the surface, was calculated and applied on the boundary particles of the droplet. We present the simulation result of droplet motion with gravity force. By using the basic method of SPH for fluid modeling, and a combination of 3D Free-Surface Detection algorithm with MLS method, we can simulate droplet phenomena with expected result.*

## 1 Introduction

Fluid simulation is a tool for generating realistic animations of water, smoke, explosions, and related phenomena that evolves the motion of the fluid forward in time by using Navier-Stokes equation, the classical Newtonian hydrodynamics which describe the physics of fluids. There are several competing techniques in computational fluid dynamics for fluid simulation and one of the most popular is Smoothed Particle Hydrodynamics (SPH) method.

The SPH is a meshfree, lagrangian method with particle-based interpolation. Fluids are modeled using particles. Every particle carries its attributes such as density, pressure, and viscosity. A lagrangian form of the Navier-Stokes equation can be used to calculate those attributes over time, finding the new acceleration and then new position for each particle.

SPH method was firstly proposed for astrophysical problems by Lucy (1977) [1] and Gingold & Monaghan (1977) [2], who used this method to study dynamical fission instabilities in rapidly rotating stars. Many cases in the field of astrophysical fluid dynamics processes have been studied numerically in 3D using SPH by Monaghan (1992) [3]. Since then, SPH has been used to study in the field of astrophysical topics such as formation of galaxies, supernova explosion, stellar collisions, formation of stars and planets, and many more. Nowadays, SPH has been applied to a wide range of problem fields, including material science, free surface flows, oceanography, volcanology and several computation fluid dynamics (CFD) applications. Some fluid dynamics applications of SPH that have been studied are: applications to heat conduction by Jeong, Jhon, Halow & van Osdol (2003) [4], solidification by Monaghan, Huppert & Worster (2005) [5], simulation of shock waves by Nejad-Asghar (2006) [6], modeling of wave interaction with porous media by Shao (2010) [7], and so on.

The main advantage of SPH is the possibility of dealing with larger local distortion than mesh-based or eulerian methods. Moreover, simulation detail and size are limited by mesh or grid resolution and size. They do not occur if we used lagrangian or meshfree methods.

In this paper, we focus to investigate and simulate liquid droplet in 3D. There are many droplet phenomena that can be found in nature such as dew or water beading on a leaf, drizzle or water dripping from a tap, ocean spray, waterfall mist and so on. In these cases, surface tension is the important aspect for the shape of fluids or droplets. In the absence of other forces, including gravity, in real phenomena the shape of a droplet for all kinds of liquid is perfectly spherical.

By SPH method, a droplet is represented by a set of particles. In order to give a surface tension to the droplet, we need to find all particles on the boundary which are interpolated to be a boundary surface by using free-surface detection algorithm. We used simplified free-surface detection based on the idea that have been proposed by Marrone, Colagrossi, Le Touze & Craziani (2010) [8]. After the boundary surface has been determined, we can include the surface tension to make the simulation approximate the real shape of droplets.

From the previous work, simulation of surface tension in 2D and 3D with SPH method by Zhang (2010) [9], the SPH boundary surface in 3D is reconstructed locally with Moving Least-Squares (MLS) method. Surface tension was modeled depending on the curvature, normal vector, particle spacing and coefficient of surface tension, where, the curvature and normal vector are calculated according to the reconstructed interface. This paper presents application of the method to simulate liquid droplet by adding artificial viscosity on the particles at boundary for modeling surface tension. The simulation was built in C++ language. The detail of the method are given in the next section.

## 2 Numerical Model

To apply the SPH particle approximation, fluids are modeled based on governing equation in non-conservation form given by the following [10]:

$$\frac{D\rho}{Dt} = -\rho \nabla \cdot \vec{v}, \quad (1)$$

$$\frac{D\vec{v}}{Dt} = -\frac{1}{\rho}\nabla p + \vec{F}, \quad (2)$$

$$p = c^2(\rho - \rho_0). \quad (3)$$

Eq. (1) is called continuity equation where $t$, $\rho$, and $\vec{v}$ represent time, density, and velocity vector, respectively. Eq. (2) is known as momentum equation where $p$ is pressure and $\vec{F}$ is external force. Pressure is calculated according to the equation of state, Eq. (3), with $c$ is sound speed and $\rho_0$ the reference density of fluid.

We use the smoothing kernel function and its gradient as follows [11]

$$W_{ij} = W(R,h) = \beta \begin{cases} \frac{2}{3} - R^2 + \frac{1}{2}R^3, & 0 \leq R < 1 \\ \frac{1}{6}(2-R)^3, & 1 \leq R < 2, \\ 0, & 2 \geq R \end{cases}$$

$$\nabla_i W_{ij} = \nabla W(R,h) = \frac{\beta}{h} \begin{cases} -2R + \frac{3}{2}R^2, & 0 \leq R < 1 \\ \frac{1}{2}(2-R)^2, & 1 \leq R < 2, \\ 0, & 2 \geq R \end{cases}$$

where in three-dimensional case,

$$\beta = \frac{3}{2}\pi h^3,$$

$$R = \frac{|\vec{r}_i - \vec{r}_j|}{h}.$$

Here, $h$ represents the smoothing length, and $\vec{r}_{i,j}$ is the position of particle $i$ or $j$. The principle behind the smoothing kernel function is to determine how much influence a particle has on other particles within the smoothing length. Monaghan & Lattanzio (1985) [12] devised this smoothing function based on the cubic spline known as the B-spline function.

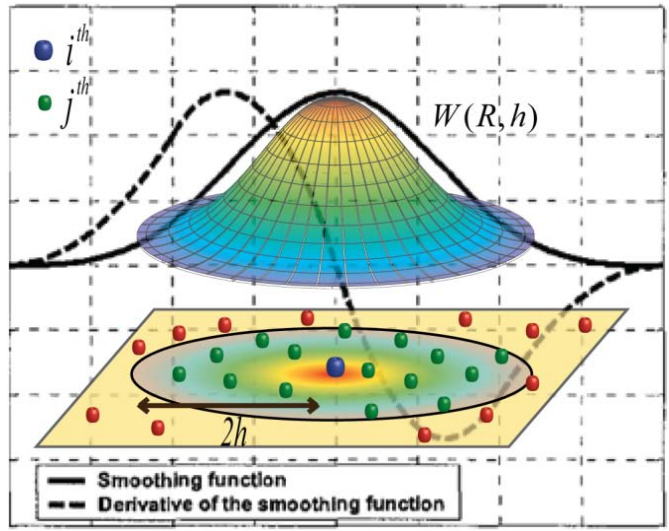

**Fig. 1.** The cubic spline kernel.

*2.1 Fluid Model*

By the discretization in the standard SPH method [11], for each particle, equation of continuity Eq. (1) can be written as follows

$$\frac{d\rho_i}{dt} = \sum_{j}^{N} m_j (\vec{v}_i - \vec{v}_j) \cdot \nabla_i W_{ij},$$

and momentum equation Eq. (2) can be discretized as below

$$\frac{d\vec{v}_i}{dt} = -\sum_j^N m_j \left( \frac{p_i + p_j}{\rho_i \rho_j} + \Pi_{ij} \right) \cdot \nabla_i W_{ij} + \vec{F},$$

where, $m$ denotes the mass of the particle, $\vec{F}$ is external force such as gravity force and momentum force from the collision with the wall, and $\Pi_{ij}$ is artificial viscosity for stabilization as given by [3]

$$\Pi_{ij} = \begin{cases} \dfrac{-\alpha c \mu_{ij} + \beta \mu_{ij}^2}{\rho_{ij}}, & (\vec{v}_i - \vec{v}_j) \cdot (\vec{r}_i - \vec{r}_j) < 0 \\ 0, & (\vec{v}_i - \vec{v}_j) \cdot (\vec{r}_i - \vec{r}_j) \geq 0 \end{cases}, \quad (4)$$

with

$$\mu_{ij} = \frac{h(\vec{r}_i - \vec{r}_j) \cdot (\vec{v}_i - \vec{v}_j)}{|\vec{r}_i - \vec{r}_j|^2 + \eta^2},$$

$$\rho_{ij} = \frac{\rho_i + \rho_j}{2}.$$

The expression for $\Pi_{ij}$ contains a term which produces an artificial shear and bulk viscosity. The artificial viscosity associated with $\alpha$ produces a bulk viscosity, while the second term associated with $\beta$ acts as shear viscosity. The factor $\eta$ is inserted to avoid numerical divergences when two particles are approaching each other [11].

The method to get the position and velocity for each particle in a timestep uses Leapfrog Time Integration according to the following scheme:

$$\vec{r}_i^* = \vec{r}_i^n + \frac{dt}{2} \vec{v}_i^n,$$

$$\rho_i^{n+1} = \rho_i^n + \frac{d\rho_i}{dt} dt,$$

$$\vec{v}_i^{n+1} = \vec{v}_i^n + \dot{\vec{v}}_i^n dt,$$

$$\vec{r}_i^{n+1} = \vec{r}_i^* + \frac{dt}{2} \vec{v}_i^{n+1},$$

where the initial position, velocity and density for each particle ($\vec{r}_i^0$, $\vec{v}_i^0$, and $\rho_i^0 = \rho_0$, respectively) are given at the beginning of the simulation.

XSPH [3] is used in the calculation and the velocity is revised by

$$\frac{d\vec{r}_i}{dt} = \vec{v}_i - \varepsilon \sum_j^N \frac{m_j}{\rho_j} (\vec{v}_i - \vec{v}_j) W_{ij},$$

where $\varepsilon$ is a constant ranging from 0 to 1. XSPH makes the SPH particles motion well-ordered.

*2.2 Surface Tension Model*

Surface tension is an effect of interparticle Van der Waals attraction in which particles at or near the surface do not have other neighbors of the same kind on all sides, while particle not near the surface are attracted to other particles in their neighborhood equally in all directions and thus the total force for particle in the interior is equal to zero. For particles at or near the surface, the total force tends to pull them into the interior. This results in the liquid having an elastic-like skin. Surface tension only works on the surface of the fluids and has the dimension of force per unit area. Curvature is a measure of how sharply a curve or surface bends, which determines the value of surface tension force.

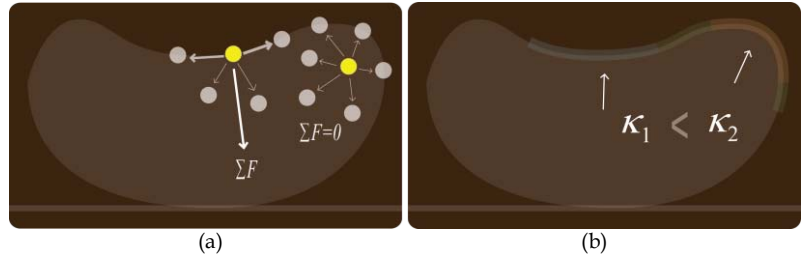

(a)  (b)

**Fig. 2.** The droplet's shape in 2D to illustrate the surface tension (a) and the curvature, $\kappa$, (b).

In order to obtain surface tension acting on the boundary surface, we need to find particles located on the boundary and then transform their coordinate into local coordinate system. After that, we can reconstruct the boundary surface using the set of local particle boundary to determine the normal vector and curvature of the surface, which are used to calculate the surface tension force in that area. Additional treatment between particles at the boundary is given in the next step. The detailed explanation for each procedure is given in the following.

### 2.2.1 Boundary Particle Detection

The surface detection algorithm uses the unique gradient field function properties near its boundary. In SPH integral approximation, gradient of the kernel should integrate to zero vector inside a fluid domain and serve as inverted surface normal vector at the fluid boundary. In SPH discretization, the above calculation is brought by a summation using gradient of the kernel. The length of this summation vector should lie near zero inside the fluid domain, and serve as approximation of inverted normal vector at the fluid boundary. Applying a threshold value to the length of this vector can be used to determine the particle's boundary status.

The determination of particle's boundary status can be enhanced by creating a scan cone around the previously calculated normal vector. This cone checks if there is any other covering particle inside the kernel domain.

### 2.2.2 Coordinate System Transformation

After the boundary particles are detected, the next step is to transform the coordinates of boundary particles into a local coordinate system. The advantage of the transformation of the coordinate system is that the boundary surface is guaranteed to be one-valued in the local coordinate system [9]. To transform the coordinate system there are 3 step:

1. Set the local origin of the local coordinate system

$$\vec{r}_{O'} = \sum_{j}^{N} \vec{r}_j / N,$$

2. Set the local three-dimensional basis vector

$$\hat{z}_{O'} = \frac{\vec{r} - \vec{r}_{O'}}{\|\vec{r} - \vec{r}_{O'}\|}, \qquad \hat{x}_{O'} = \begin{cases} \dfrac{\hat{x} \times \hat{z}_{O'}}{\|\hat{x} \times \hat{z}_{O'}\|}, & \hat{y} \times \hat{z}_{O'} = 0 \\ \dfrac{\hat{y} \times \hat{z}_{O'}}{\|\hat{y} \times \hat{z}_{O'}\|}, & \hat{y} \times \hat{z}_{O'} \neq 0 \end{cases}, \qquad \hat{y}_{O'} = \frac{\hat{z}_{O'} \times \hat{x}_{O'}}{\|\hat{z}_{O'} \times \hat{x}_{O'}\|},$$

3. Transform the coordinate into local coordinate

$$x' = \vec{r}_O \cdot \hat{x}_{O'},$$
$$y' = \vec{r}_O \cdot \hat{y}_{O'},$$
$$z' = \vec{r}_O \cdot \hat{z}_{O'}.$$

*2.2.3 Surface Reconstruction*

The boundary particle coordinates are transformed into local coordinate system, and then the boundary surface can be reconstructed by using Moving Least-Squares (MLS) method.

MLS is a method to reconstruct the surface from a set of points, by adding weighted function in the region around the point of which the reconstructed value is requested. We need to get the boundary surface of a function given as below

$$f(x', y') = c_0 + c_1 x' + c_2 y' + c_3 x'^2 + c_4 x' y' + c_5 y'^2.$$

For each particle, the coefficients $c_0, c_1, c_2, c_3, c_4,$ and $c_5$ can be obtained by minimizing

$$R^2 = \sum_j^N W_{ij} [z'_j - f(x'_j, y'_j)]^2 = \sum_j^N W_{ij} [z'_j - (c_0 + c_1 x'_j + c_2 y'_j + c_3 x'^2_j + c_4 x'_j y'_j + c_5 y'^2_j)]^2.$$

In other words,

$$\frac{\partial R^2}{\partial c_0} = -2 \sum_j^N W_{ij} [z'_j - (c_0 + c_1 x'_j + c_2 y'_j + c_3 x'^2_j + c_4 x'_j y'_j + c_5 y'^2_j)] = 0,$$

$$\frac{\partial R^2}{\partial c_1} = -2 \sum_j^N W_{ij} [z'_j - (c_0 + c_1 x'_j + c_2 y'_j + c_3 x'^2_j + c_4 x'_j y'_j + c_5 y'^2_j)] x'_j = 0,$$

$$\vdots$$

$$\frac{\partial R^2}{\partial c_5} = -2 \sum_j^N W_{ij} [z'_j - (c_0 + c_1 x'_j + c_2 y'_j + c_3 x'^2_j + c_4 x'_j y'_j + c_5 y'^2_j)] y'^2_j = 0.$$

These lead to the system of equations

$$c_0 \sum_j^N W_{ij} + c_1 \sum_j^N W_{ij} x'_j + c_2 \sum_j^N W_{ij} y'_j + c_3 \sum_j^N W_{ij} x'^2_j + c_4 \sum_j^N W_{ij} x'_j y'_j + c_5 \sum_j^N W_{ij} y'^2_j = \sum_j^N W_{ij} z'_j,$$

$$c_0 \sum_j^N W_{ij} x'_j + c_1 \sum_j^N W_{ij} x'^2_j + c_2 \sum_j^N W_{ij} x'_j y'_j + c_3 \sum_j^N W_{ij} x'^3_j + c_4 \sum_j^N W_{ij} x'^2_j y'_j + c_5 \sum_j^N W_{ij} x'_j y'^2_j = \sum_j^N W_{ij} z'_j x'_j,$$

$$\vdots$$

$$c_0 \sum_j^N W_{ij} y'^2_j + c_1 \sum_j^N W_{ij} x'_j y'^2_j + c_2 \sum_j^N W_{ij} y'^3_j + c_3 \sum_j^N W_{ij} x'^2_j y'^2_j + c_4 \sum_j^N W_{ij} x'_j y'^3_j + c_5 \sum_j^N W_{ij} y'^4_j = \sum_j^N W_{ij} z'_j y'^2_j.$$

Therefore, the coefficients of surface function can be written in matrix form:

$$\mathbf{C} = \begin{bmatrix} c_0 \\ c_1 \\ c_2 \\ c_3 \\ c_4 \\ c_5 \end{bmatrix} = \left[\sum_{j}^{N} W_{ij} \mathbf{P}_j \mathbf{P}_j^t \right]^{-1} \left[\sum_{j}^{N} W_{ij} z'_j \mathbf{P}_j \right],$$

where $\mathbf{P}_j$ is basis quadratic polynomial matrix for each surface function interpolation

$$\mathbf{P}_j = \begin{bmatrix} 1 \\ x'_j \\ y'_j \\ x'^2_j \\ x'_j y'_j \\ y'^2_j \end{bmatrix}.$$

*2.2.4 Surface Tension Force*

After the surface fuction is obtained, we can find the normal vector of the surface by the formula

$$\vec{n} = \frac{1}{\sqrt{1 + f_x^2 + f_y^2}} \langle -f_x, -f_x, 1 \rangle,$$

and the curvature is given as

$$\kappa = \frac{(1 + f_x^2) f_{yy} - 2 f_x f_y f_{xy} + (1 + f_y^2) f_{xx}}{2(1 + f_x^2 + f_y^2)^{3/2}},$$

where

$$f_x = \frac{df(x,y)}{dx},\ f_y = \frac{df(x,y)}{dy},\ f_{xx} = \frac{d^2 f(x,y)}{dx^2},\ f_{yy} = \frac{d^2 f(x,y)}{dy^2},\ f_{xy} = \frac{d^2 f(x,y)}{dxdy}.$$

The surface tension force is calculated as below [9]

$$\vec{f} = \frac{\sigma \vec{n} \kappa}{\varepsilon},$$

where $\sigma$ is the coefficient of surface tension, $\vec{n}$ is normal vector of surface, $\kappa$ is the curvature of the surface, and $\varepsilon$ takes the value of the particle spacing.

*2.2.5 Overcoming the Unstability of the Boundary Particle*

The presence of surface tension force to the particles at the boundary, and the pressure of the particles at the interior, make movement of particles on the surface unstable if we rely only on surface tension force. To overcome this, we increase the coefficients $\alpha$ and $\beta$ from Eq. (4) only for particles on the surface, so that movement of particles on the surface is more well-organized.

## 3 Results and Discussion

Deformation of a 3D liquid drop from initial shape of a cube to a sphere is simulated. The side length of the cube is 0.03, the particle spacing is 0.002, and 3375 SPH particles are used in the simulation. The coefficient of surface tension $\sigma$ is set to 500. The $\alpha$ and $\beta$ for particles at the interior are 0.3 and 0.0, and for particles at the boundary are 1.0 and 1.0, respectively.

**Fig. 3** shows snapshots of falling cubic liquid droplet with colors describing the magnitude of the curvature. Due to surface tension force, the eight corners of liquid droplet which have larger curvature deform the droplet into a shape like a diamond at $t = 0.04$ until it becomes almost sphere before hitting the floor. After the droplet hits the floor at $t = 1.0$ (see **Fig. 4**), the droplet tries to keep the shape of a spherical cap.

However, unnatural impact happens because of the difference on $\alpha$ and $\beta$ between boundary and interior particles. The center of mass of the droplet is shiffting on horizontal plane during the simulation. To overcome this problem, our idea is to set the center of mass to be unaffected by any unbalance total force of the surface tension. This idea is currently under investigation.

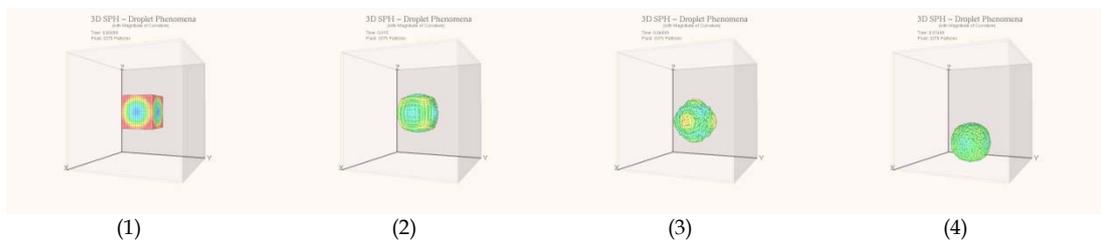

 (1)    (2)    (3)    (4)

**Fig. 3.** Snapshots of 3D SPH for droplet phenomena with gravity force before hitting the floor.

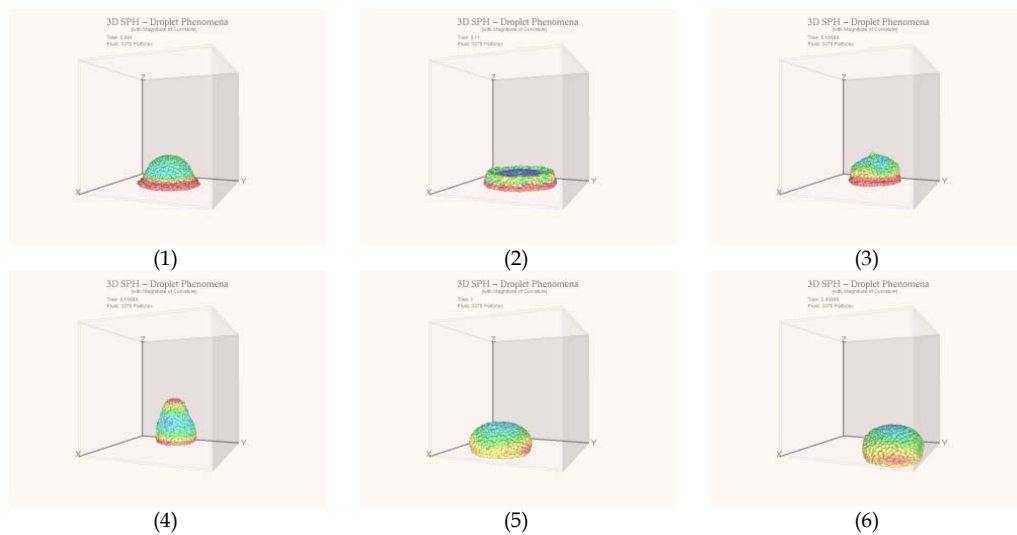

 (1)    (2)    (3)

 (4)    (5)    (6)

**Fig. 4.** Snapshots of 3D SPH for droplet phenomena with gravity force after hitting the floor.

## 4   Conclusion

The SPH method has been successfully applied to simulate droplet phenomena. As we know, surface tension is important in the case of liquid droplet. Since the boundary particles are detected explicitly, the surface can be reconstructed by using MLS and surface tension can be calculated directly on the surface. Surface tension force acts only on the surface which depends on the curvature. In order to reduce the effects of particle deficiency on the surface, adding a treatment like increasing the coefficients of bulk and shear for artificial viscosity only for particles at the boundary is successful in making the particle on surface move regularly. The result showed that the droplet tries to keep its shape of a sphere as expected.